# Size matters: performance declines if your pixels are too big or too small

Vassilis Kostakos and Eamonn O'Neill
Department of Computer Science
University of Bath, UK
{vk, eamonn}@cs.bath.ac.uk

## Abstract

We present a conceptual model that describes the effect of pixel size on target acquisition. We demonstrate the use of our conceptual model by applying it to predict and explain the results of an experiment to evaluate users' performance in a target acquisition task involving three distinct display sizes: standard desktop, small and large displays. The results indicate that users are fastest on standard desktop displays, undershoots are the most common error on small displays and overshoots are the most common error on large displays. We propose heuristics to maintain usability when changing displays. Finally, we contribute to the growing body of evidence that amplitude does affect performance in a display-based pointing task.

**CR Categories:** H.5.2 [Information Interfaces and Presentation]: User Interfaces – Graphical user interfaces, Input devices and strategies, Interaction styles, Theory and methods.

## 1 Introduction

Despite extensive ongoing research into multimodal interaction [e.g. Grawala et al., 1997] graphical displays and visual feedback remain central to our use of computing systems. Within pervasive or ubiquitous computing, the use of diverse input and output devices is characterized by displays of various form factors, ranging from mobile phones and PDAs to video walls and projectors.

An ongoing research and design challenge is the development and specialisation of interfaces for this range of different display sizes [e.g. Wei et al., 2000]. Consider the familiar example of a user connecting her notebook computer to an external display or projector that simply scales up or scales down the contents of the notebook's display (Figure 1). How does this scaling affect usability? To answer this question, we first need to understand the role of display size, and its relation to other parameters such as display resolution, mouse sensitivity, mouse movement and target size.

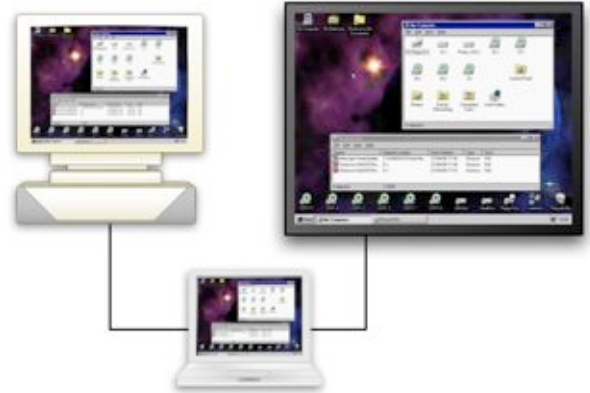

**Figure 1.** Varying display sizes being used for the same interface and contents.

In this paper, we extend an existing conceptual model of the relationship between mouse input and visual output [Blanch et al., 2005]. Typically, only mouse sensitivity has been considered as the main parameter in such models [e.g. Accot & Zhai, 2001]. We argue that the relationship between mouse and pointer movement is affected by two independent parameters: mouse sensitivity and pixel size (i.e. how much space a pixel takes up in the real world).

Furthermore, we distinguish between two different types of "scale". First, scale can relate to how far participants need to move their hands (for example, in cm) in order to reach a target. In another sense, scale can relate to how far (in cm) the mouse pointer travels on the display.

Having developed our model, we then proceed to test its validity. We report a controlled experiment in which mouse sensitivity was kept constant and only pixel size was varied by using three different displays. We evaluated participants' performance on target acquisition tasks using Fitts' law [Fitts, 1954]. Using our model we are able to describe the differences in performance across the experimental conditions. Furthermore, our results show that the two types of scale have independent effects on performance.

In the following section we provide a brief overview of Fitts' law and relevant work, and then introduce our conceptual model for mapping mouse movement to pointer movement. Subsequently, we describe our experiment and the results we obtained, and conclude with a discussion of our results and future work.

## 2 Related work

Fitts' law is a predictive and descriptive tool that relates the time required for target acquisition to characteristics of the target. It has gone through a number of reformulations, and currently the most accepted one, established in ISO 9241-9, is the Shannon formulation

$$MT = a + b \cdot ID \quad ID = \log_2\left(\frac{A}{W} + 1\right)$$

where MT is the time taken to reach a target, $a$ and $b$ are empirically derived constants, ID is the index of difficulty, A is the amplitude (distance) between targets and W is the width of targets (see Figure 2).

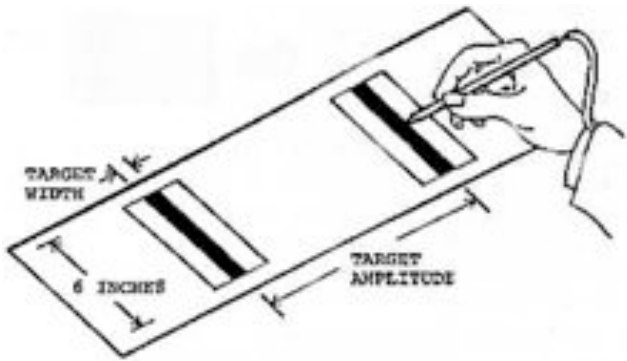

**Figure 2.** Fitts' original experiment.

Fitts' law has been applied in human-computer interaction (HCI) studies in various ways and forms, in some cases with debatable appropriateness. In our experimental design we draw on existing work that documents appropriate ways of using Fitts' law in HCI.

For instance, Soukoreff & MacKenzie [2004] provide seven suggestions for carrying out consistent studies using Fitts' law, which relate to the formulation and range of ID, and recording and adjusting for errors. By drawing on previous studies, they demonstrate that following these suggestions results in consistent findings.

Furthermore, Guiard [2001] criticizes the practice of using amplitude and width as independent variables in studies, and suggests that they are in fact confounding variables. He proposes the use of amplitude and ID as independent variables.

Ultimately, studies aim to provide a performance index for each device or interface, which can be meaningful outside the experimental conditions. This index is known as throughput. Conventionally, throughput has been calculated as *MT/ID*. However, Zhai [2004] has argued that this is an ill-defined concept that cannot be generalized beyond experimental conditions. He suggests the use of 1/*b* as the determinant of throughput.

In specifying the desirable properties of their model that extends Fitts' Law's one-dimensional model to bivariate pointing, Accot and Zhai [2003] include scale independency; that is, multiplying amplitude and width (and, in their study, height of target) by a constant leaves movement time unchanged. Although scale independency is a property of Fitts' original equation, the Welford formulation and even the Shannon formulation, there is evidence that it does not model what actually happens in a Fitts' law pointing task.

Gan and Hoffmann [1988] demonstrate that in an original Fitts' pointing task (that is using two metal plates and a stylus) "regardless of ID, there was a significant effect of amplitude on movement times". Guiard [2001] notes that although "Fitts' law predicts that MT […] should be scale independent […] there is ample evidence that, in violation of the law, scale [i.e. amplitude] affects MT quite substantially". Although amplitude has been shown to affect performance [Blanch et al., 2004; Guiard & Beaudouin-Lafon, 2004], there is still an ongoing debate about this issue.

### 2.1 Mouse and pointer movement

There exists work that addresses mouse and pointer movement in relation to Fitts' law. For instance, Accot & Zhai [2001] have studied the effect of display scale in steering tasks, in an experiment involving a graphics tablet and a constant display size and resolution. Here the authors vary pointer sensitivity in order to study the performance of different muscles and limbs when users moved their hands.

In another study, Guiard & Beaudouin-Lafon [2004] examine the effects of zooming techniques (bi-focal, fish-eye, pan-and-zoom) on target acquisition and interface navigation. In their study they do not alter display size, but instead consider scale in terms of zooming in and out of the contents of the display. This is essentially equivalent to dynamically changing the display resolution in certain parts of the display.

Furthermore, Blanch et al. [2004] demonstrate a performance improvement by providing dynamic and adaptable pointer sensitivity. Specifically, they show a reduction of 1 bit in ID for every doubling of pointer sensitivity.

## 3 A conceptual model

Traditionally, distance and scale have been considered in relation to the amplitude between targets. An increase in "scale" has been instantiated as an increase of the amplitude between targets. The resulting effect is an increase of the physical movement required to reach the targets.

Yet, in Fitts' original experiment [Fitts, 1954], distance was a measure of how much participants had to move their

hand, not how far apart the targets were. It so happened that there was a one-to-one mapping between

- the distance that participant's *hand* moved,
- the distance the *stylus* moved, and
- the distance between *targets*.

With the introduction of computers' graphical displays and mice, however, this one-to-one mapping between hand movement, pointer movement and target distance was lost. Moving the mouse by 5 cm could move the cursor by anything from 50 to 500 pixels, depending on mouse sensitivity. Furthermore, 500 pixels could appear to be the length of a credit card or the length of an A4 sheet of paper, depending on pixel size. Pixel size in turn is determined by display size and display resolution. Increasing a display's resolution results in smaller pixels, while decreasing the resolution results in larger pixels. Similarly, using a larger display to present the same number of pixels results in larger pixels, while the opposite is true of a smaller display.

Hence, when using a computer display and mouse (but not, it should be noted, when using a stylus directly on a touchscreen display), an increase in "scale" can be considered as *either* an increase in the distance that the user's hand needs to move *or* the distance that the mouse pointer travels across the display.

To disentangle these concepts, we extend the model presented in [Blanch et al., 2004]. Here, we discuss the notions of

- *motor* space: how much our hands move in relation to the physical environment (i.e. how many meters)
- *virtual* space: how much the pointer moves in relation to other graphical objects on the display (i.e. how many pixels)
- *display* space: the size of graphical display objects in relation to the physical world (i.e. how many meters)

The top half of Figure 3 represents the mapping between the distance that a user's hand moves (motor space) and the pixel distance that the pointer on the display is moved (virtual space). The line labeled "pointer sensitivity" represents this mapping. As this mapping becomes vertical, the physical movement needed to acquire a target increases and the pointer appears to move slowly in relation to other objects on the display. As the mapping becomes horizontal, the physical movement to acquire a target decreases and the pointer appears to move fast in relation to other objects on the display.

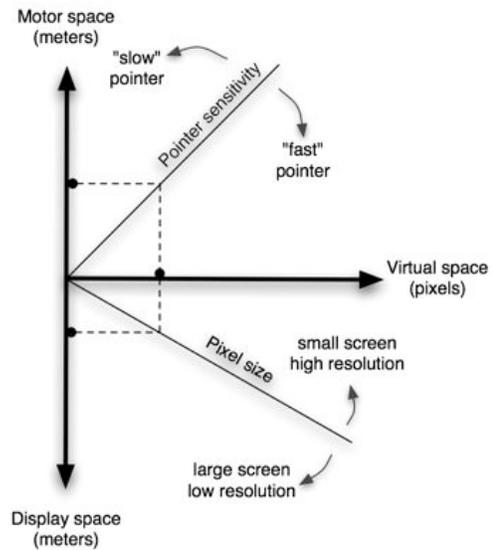

**Figure 3.** The mapping between motor space and virtual space is given by pointer sensitivity. The mapping between virtual space and display space is given by pixel size.

Our extension to Blanch et al's [2004] model consists in the lower half of Figure 3. The key idea is that both display size and resolution affect the size of pixels in relation to the physical world. The lower half of Figure 3 represents the mapping between digital pixels (virtual space) and the actual rendering of those pixels on the surface of the display (display space). The line labeled "pixel size" represents this mapping. As this mapping becomes horizontal, the pixels take up less space in the physical world. In this case, objects on the display appear to move slowly in relation to the *physical world*. This is a characteristic of displays with a high density of pixels (high resolution), and is typical of smaller displays.

As the pixel size mapping becomes vertical, the pixels take up more space in the physical world. This is a characteristic of displays with a low density of pixels (low resolution), and is typical of larger displays and video projectors. In this case, objects on the display appear to move quickly in relation to the *physical world*.

## 3.1 Implications of our model

Previous research has assumed constant display size and pixel size, effectively focusing on what we represent in the top half of our model. These assumptions imply that only mouse sensitivity can have an effect on Control-Display ratio (C-D ratio), i.e. how motor space movement is transformed to display space movement. Thus, conventionally, the C-D ratio has been seen purely as a function of pointer sensitivity [Blanch et al., 2004], while we maintain that C-D ratio is a function of both pointer sensitivity and pixel size.

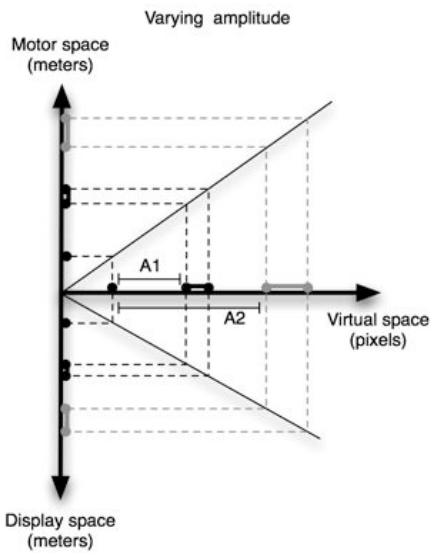 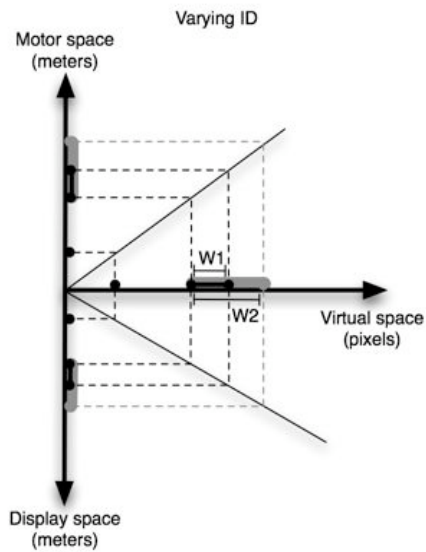 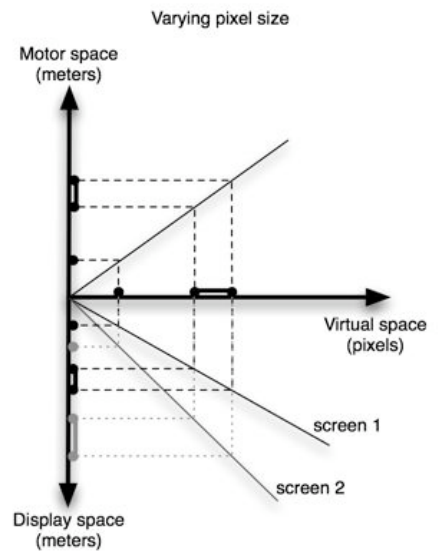

**Figure 4.** Conceptualizing a variation in amplitude. Note that width changes to keep ID constant.

**Figure 5.** Conceptualizing a variation in ID. Note that amplitude remains constant.

**Figure 6.** Conceptualizing a variation in pixel size (by changing display size or resolution).

To test the effects of pixel size on target acquisition we designed a controlled experiment where participants carried out a number of target acquisition trials on three different display sizes. By using three distinct display sizes, we effectively varied pixel size while keeping mouse sensitivity constant.

As described above, changing the display size is one way of changing pixel size, while changing the resolution of the display is another (Figure 3). In our experiment, we changed only display size in order to manipulate pixel size. Differences in the displays' native resolutions and pixel densities were accounted for by physically measuring pixel sizes and comparing them across the displays.

By varying the number of pixels between targets (amplitude) we varied how much participants needed to move their hands. Because we kept mouse sensitivity and gain function constant across the conditions, an increase in the amplitude between two targets would increase the motor distance between those targets by the same amount in each condition.

In addition to varying *pixel size* and *amplitude*, we also varied *ID*, which is the actual difficulty of task as predicted by Fitts' law. The conceptual differences of altering each of these three variables are shown in terms of our model in Figures 4, 5 and 6.

In Figure 4 we see that we can vary amplitude while keeping ID and pixel size constant. We achieve this by proportionately changing the pixel width of targets. In Figure 5 we vary ID by keeping amplitude constant and varying only the pixel width of targets. In Figure 6, we vary only pixel size while amplitude and ID remain constant. This is equivalent to simply plugging in a different display without changing anything else.

It is important to note that our experiment manipulated only pixel size and not viewing angle. Viewing angle is a function of two variables: pixel size and viewing distance. Since we kept viewing distance constant and varied only pixel size, viewing angle changed in proportion to pixel size. Specifically, it is not the case that viewing angle affects pixel size, but rather pixel size affects viewing angle.

## 4 The Experiment

### 4.1 Design

Our experiment was within subjects, and we used a three-way (3x3x7) related samples factorial design. The independent variables were Screen size (small, medium and large), Amplitude (200, 400 and 600 pixels) and Index of Difficulty (2.58, 2.94, 3.46, 3.75, 3.93, 4.14, and 4.39 bits). We measured movement time (milliseconds) and number of errors (overshoots, undershoots and other errors).

Order and carry-over effects were controlled by asking the participants to take short breaks in a different room. We dealt with any remaining order, carry-over or learning effects by counter-balancing Screen size and randomizing Amplitude and ID.

It was predicted that (i) movement time is longer with large displays than with small displays, and (ii) movement time is affected by amplitude independently of ID.

### 4.2 Participants

We had 60 participants, 38 male and 22 female. The participants ranged in age from 18 to 23. Participants were allocated randomly to each of the 63 conditions, and each participant used her preferred hand.

### 4.3 Apparatus

The experiment was conducted on a Xybernaut wearable PC (500 MHz Intel Mobile Celeron). This unit ran Microsoft Windows 2000 and had an ATI Rage Mobility-M Graphics card. The mouse sensitivity was set to default and remained constant throughout the experiment. The input device was a Microsoft IntelliMouse Optical USB. The Xybernaut PC was chosen because it could drive the range of display sizes we were interested in while keeping the remainder of the computing equipment constant.

The PC was connected to one of our three displays for each condition. The conditions were performed in full screen mode, with a constant background color. The three displays used were:

- Small: Xybernaut display. (8.4", 173x131mm, all-light readable display, 800 x 600 color SVGA, FPD-00200, **pixel size**: 0.047 mm$^2$).
- Medium: Ilyama display. (18.1", 362x290mm, 1280 x 1024, AS4612UT BK, **pixel size**: 0.291 mm$^2$).
- Large: NEC plasma display. (61", 1351 x 768 mm, 1014x768mm used, **pixel size**: 2.162 mm$^2$).

The medium display was 2.1 times wider than the small display. The large display was 2.8 times wider than the medium display. Additionally, pixel size increased across the three displays in the ratio 0.1615:1:7.4296.

Participants sat directly in front of the displays at a constant distance from the displays. The centre points of the displays were aligned at the same height. The output from the PC was constant at 800x600 pixels, and the refresh rate was constant at 60Hz. Thus, changing the display size directly manipulated pixel size. Table 1 shows the variation in target size related to Amplitude and ID. Here we use the Shannon formulation of Fitts' law.

| Amplitude (pixels) | ID (bits) | | | | | | |
|---|---|---|---|---|---|---|---|
| | 4.39 | 4.14 | 3.93 | 3.75 | 3.46 | 2.94 | 2.58 |
| 200 | 10 | 12 | 14 | 16 | 20 | 30 | 40 |
| 400 | 20 | 24 | 28 | 32 | 40 | 60 | 80 |
| 600 | 30 | 36 | 41 | 48 | 60 | 90 | 120 |

**Table 1**. Target sizes (in pixels) in virtual space.

| Screen Size | Amplitude (horizontal) | | |
|---|---|---|---|
| | 200px | 400px | 600px |
| Small | 4.33 | 8.65 | 12.98 |
| Medium | 9.05 | 18.10 | 27.15 |
| Large | 23.35 | 50.70 | 76.05 |

**Table 2**. Display space representation of amplitudes (in cm). For example, if we used a ruler to measure the length of 400 pixels on the medium display we would find a length of 18.10 cm.

In Table 2 we show the display space representation of the different amplitudes in each of our three display conditions. Note again that display size had no relationship with the actual distance that the mouse had to move in order for participants to click the targets.

### 4.4 Procedure

The experiment included two sessions for each display: a practice session and a data-collection session. The practice session lasted until participants reached an acceptable level of errors (fewer than 4 per condition). The data-collection session consisted of participants testing the 63 conditions for Screen size, Amplitude and ID. Within each condition, participants performed 10 trials.

Participants were instructed to sit and read the instructions that appeared on screen. They first completed a practice session and were instructed to begin whenever they were ready. The instructions specified that the participant should sit in an upright position to keep the distance from the displays constant, and place the mouse mat in the most comfortable position on the desk. Participants were reminded to sit upright if they leant forward or backward.

The task was to point to two square targets on the display, shown in Figure 7.

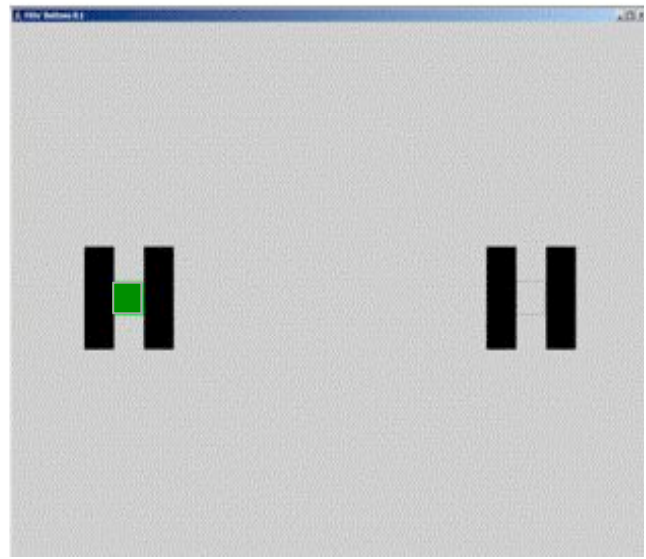

**Figure 7**. The green target indicates the active target (in this case on the left). The vertical bars indicate the overshoot and undershoot areas, and are highlighted here for illustrative purposes. They were not visible in the experiment.

The active target was green, the passive target grey. The instructions requested participants to "click and release the mouse on the GREEN button as quickly and as accurately as possible". When participants clicked on the highlighted target, or missed, the colors swapped. When the colors swapped, this indicated that the clicking task was completed and that the participant was to click on the opposite target.

Clicking off the target recorded an error. We measured errors by counting the number of clicks in four areas around the target (on either side of the target to measure overshoot and undershoot, above the target and below the target). When clicked on, these areas changed from light to dark grey indicating an error.

When each participant had completed the task on a display they were instructed to sit in a different room. They remained there while the next display was set up. This break took three to four minutes.

## 5 Results

For each participant we measured movement time and error. The movement times of erroneous trials were discarded. The mean was then calculated and movement time data from outliers (more than two standard deviations away from the mean) was discarded. Such data was generated when participants paused in order to reposition themselves or adjust the position of the mouse. The mean was then recalculated and this value taken as the movement time for the condition.

The data were analyzed using a three-way analysis of variance (ANOVA) for related samples, with Screen size, Amplitude and ID as the independent variables.

### 5.1 Movement time

The mean movement times are shown in Table 3.

| *Screen:* | Small | | | Medium | | | Large | | |
|---|---|---|---|---|---|---|---|---|---|
| *Amplitude:* | 200 | 400 | 600 | 200 | 400 | 600 | 200 | 400 | 600 |
| *ID* | | | | | | | | | |
| 4.39 | 951 | 828 | 809 | 872 | 747 | 758 | 937 | 820 | 808 |
| 4.14 | 884 | 763 | 767 | 805 | 732 | 723 | 852 | 784 | 786 |
| 3.93 | 820 | 742 | 735 | 767 | 709 | 686 | 825 | 741 | 760 |
| 3.75 | 790 | 705 | 726 | 738 | 654 | 670 | 757 | 711 | 727 |
| 3.46 | 723 | 657 | 673 | 681 | 628 | 645 | 716 | 676 | 694 |
| 2.94 | 635 | 588 | 608 | 615 | 573 | 570 | 654 | 624 | 631 |
| 2.58 | 574 | 560 | 577 | 567 | 531 | 555 | 606 | 574 | 609 |

**Table 3**. Mean movement times per condition (in milliseconds).

There was a significant effect of

- Screen size ($F_{2,118}=58.73, p<.001$)
- Amplitude ($F_{1.62,118} = 171.18, p<.001$)
- ID ($F_{3.64,214.93} = 879.94, p<.001$)
- Screen size x ID ($F_{9.34,550.88} = 4.05, p < .001$)
- Amplitude x ID ($F_{8.62,508.70} = 17.33, p < .001$)

The effect of Screen size x Amplitude and of Screen size x Amplitude x ID was not significant.

A *post hoc* pairwise Bonferroni comparison was used for the main effect of Screen size. This revealed a significant main effect of Screen size between small and medium Screen size ($p < .001$), and between large and medium Screen size ($p < .001$).

Similarly, the significant main effect of Amplitude reflected a significant difference between 200 and 600 pixel Amplitude ($p < .001$), 200px and 400 pixel Amplitude ($p < .001$) and 600px and 400 pixel Amplitude ($p < .05$).

Finally, there was a significant difference between all pairwise level comparisons of ID ($p < .001$).

### 5.2 Errors

In addition to recording total errors, we distinguished between undershoot and overshoot errors. Undershoots were recorded when participants clicked before reaching the target, while overshoots were recorded when participants clicked beyond the target.

#### 5.2.1 Total errors

The mean error rates are shown in Table 4.

| *Screen:* | Small | | | Medium | | | Large | | |
|---|---|---|---|---|---|---|---|---|---|
| *Amplitude:* | 200 | 400 | 600 | 200 | 400 | 600 | 200 | 400 | 600 |
| *ID* | | | | | | | | | |
| 4.39 | 12.50 | 5.50 | 6.17 | 7.50 | 4.33 | 5.00 | 8.33 | 6.33 | 3.33 |
| 4.14 | 8.33 | 5.50 | 5.17 | 6.17 | 5.83 | 6.00 | 8.00 | 3.83 | 5.67 |
| 3.93 | 7.33 | 5.00 | 6.33 | 7.50 | 5.83 | 5.50 | 8.67 | 4.50 | 3.50 |
| 3.75 | 6.00 | 4.83 | 7.17 | 6.50 | 3.33 | 5.67 | 5.83 | 2.33 | 5.67 |
| 3.46 | 6.17 | 4.17 | 7.83 | 4.67 | 4.17 | 4.83 | 3.00 | 4.67 | 3.33 |
| 2.94 | 5.17 | 5.67 | 6.50 | 5.50 | 5.33 | 3.17 | 4.00 | 4.33 | 3.67 |
| 2.58 | 4.83 | 5.50 | 2.17 | 4.67 | 4.50 | 1.50 | 4.00 | 4.00 | 1.33 |

**Table 4**. Total errors per condition (percentage).

There was no significant effect of Screen size x Amplitude, Screen size x ID or Screen size x Amplitude x ID.

A *post hoc* pairwise Bonferroni comparison showed a significant difference between large and small Screen size ($p < .001$), while comparison between other Screen sizes did not yield significant results.

Additionally, there was a significant difference between 200 and 400 pixels Amplitudes ($p < .001$), and between 200 and 600 pixel Amplitudes ($p < .001$). There was no significant difference between 400 and 600 pixel Amplitudes.

The three-way ANOVA of errors showed that there was a significant effect of

- Screen size ($F_{(2,118)} = 9.51, p < .001$)
- Amplitude ($F_{2,118} = 20.37, p < .001$)
- ID ($F_{6,354} = 11.59, p < .001$)
- Amplitude x ID ($F_{8.39,495.01} = 4.78, p < .001$)

Finally, 9 of the 21 pairwise comparisons for ID were significant ($p < 0.05$).

### 5.2.2 Undershoot errors

The mean undershoot error rates are shown in Table 5.

| Screen: | Small | | | Medium | | | Large | | |
|---|---|---|---|---|---|---|---|---|---|
| Amplitude: | 200 | 400 | 600 | 200 | 400 | 600 | 200 | 400 | 600 |
| ID | | | | | | | | | |
| 4.39 | 3.17 | 2.83 | 3.67 | 2.83 | 1.50 | 2.00 | 3.83 | 2.17 | 1.67 |
| 4.14 | 2.50 | 2.50 | 4.17 | 2.33 | 2.17 | 2.83 | 2.67 | 0.50 | 1.17 |
| 3.93 | 2.33 | 3.00 | 4.17 | 2.33 | 2.83 | 3.00 | 2.67 | 1.17 | 1.00 |
| 3.75 | 2.17 | 2.67 | 4.83 | 1.50 | 1.50 | 2.50 | 1.00 | 0.50 | 1.83 |
| 3.46 | 2.33 | 2.33 | 5.00 | 2.00 | 2.33 | 2.33 | 0.50 | 1.67 | 1.33 |
| 2.94 | 2.33 | 3.33 | 5.00 | 1.17 | 2.33 | 1.17 | 1.17 | 0.83 | 0.67 |
| 2.58 | 1.67 | 4.33 | 2.00 | 1.67 | 2.17 | 1.00 | 0.67 | 0.67 | 1.17 |

**Table 5**. Undershoot errors per condition (percentage).

The three-way ANOVA of undershoot errors showed that there was a significant effect of

- Screen size ($F_{1.78, 105.26}$ = 26.27, $p < .001$)
- Amplitude ($F_{1.83, 108.03}$ = 3.36, $p < .05$)
- ID ($F_{6, 354}$ = 2.96, $p < .01$)
- Screen size x Amplitude ($F_{4, 236}$ = 7.75, $p < .001$)
- Amplitude x ID ($F_{8.97, 529.16}$ = 2.19, $p < .05$)

There was no significant effect of Screen size x ID or Screen size x Amplitude x ID.

A *post hoc* pairwise Bonferroni comparison showed a significant difference between all Screen sizes, small and medium ($p < .001$), medium and large ($p < .01$), and small and large ($p < .001$).

The pairwise comparisons of Amplitude showed a significant difference only between the 200 and 600 pixel Amplitude ($p < .05$).

Finally, 1 of the 21 pairwise comparisons for ID was significant ($p < 0.05$).

### 5.2.3 Overshoot errors

The mean overshoot error rates are shown in Table 6.

| Screen: | Small | | | Medium | | | Large | | |
|---|---|---|---|---|---|---|---|---|---|
| Amplitude: | 200 | 400 | 600 | 200 | 400 | 600 | 200 | 400 | 600 |
| ID | | | | | | | | | |
| 4.39 | 7.17 | 1.83 | 1.83 | 3.00 | 2.50 | 3.00 | 4.00 | 3.33 | 1.50 |
| 4.14 | 4.00 | 2.67 | 0.50 | 3.67 | 2.83 | 2.67 | 4.67 | 3.17 | 3.33 |
| 3.93 | 3.67 | 2.00 | 2.00 | 4.50 | 2.33 | 2.17 | 4.83 | 3.33 | 2.50 |
| 3.75 | 3.33 | 2.17 | 1.83 | 4.50 | 1.67 | 3.17 | 4.33 | 1.67 | 3.33 |
| 3.46 | 3.33 | 1.50 | 2.33 | 2.50 | 1.67 | 2.33 | 2.33 | 2.83 | 1.67 |
| 2.94 | 2.50 | 2.00 | 1.00 | 4.00 | 2.67 | 1.33 | 2.50 | 3.50 | 2.50 |
| 2.58 | 2.67 | 1.00 | 0.00 | 2.67 | 2.17 | 0.00 | 3.33 | 3.17 | 0.00 |

**Table 6**. Overshoot errors per condition (percentage).

The three-way ANOVA of overshoot errors showed that there was a significant effect of

- Screen size ($F_{2, 118}$ = 3.16, $p < .05$)
- Amplitude ($F_{1.17, 101.05}$ = 42.17, $p < .001$)
- ID ($F_{6, 354}$ = 6.82, $p < .001$)
- Screen size x Amplitude ($F_{4, 236}$ = 3.37, $p < .01$)
- Amplitude x ID ($F_{7.76, 457.8}$ = 2.43, $p < .05$)
- Screen size x Amplitude x ID ($F_{14.38, 848.47}$ = 1.79, $p < .05$)

There was no significant main effect of Screen size x ID.

A *post hoc* pairwise Bonferroni comparison showed a significant difference only between the small and large Screen size ($p < .05$).

There was also a significant difference between 200 and 400 pixels Amplitudes ($p < .001$), between 200 and 600 pixels Amplitudes ($p < .001$), and between 400 and 600 pixels Amplitudes ($p < .05$).

Finally, 5 of the 21 pairwise comparisons for ID were significant ($p < 0.05$).

## 5.3 Throughput

Throughput and efficiency for each display condition are shown in Table 7. We calculated participants' throughput for each display. Throughput was calculated as $1/b$ where $MT = a + b \cdot ID$ [Zhai 2004]. We used a linear regression model to calculate these values. Additionally, we calculated efficiency as correct trials per minute.

| Screen | $R^2$ | Regression Coefficients | | Throughput | Efficiency |
|---|---|---|---|---|---|
| | | a (ms) | b (ms/bit) | 1/b (bit/second) | correct per minute |
| Small | .396 | 187.7 | 144.5 | 6.92 | 77.06 |
| Medium | .39 | 196.7 | 133.4 | 7.50 | 82.72 |
| Large | .421 | 221.0 | 140.8 | 7.10 | 77.31 |

**Table 7**. The regression coefficients, throughput and efficiency for each of the three display conditions.

## 6 Discussion

We first note that all three independent variables had a significant main effect on both movement time and error. Thus, display size (and hence pixel size) did affect performance. Furthermore, our results show that, contrary to Fitts' law, amplitude has an effect on movement time. We return to this issue towards the end of our discussion.

### 6.1 Movement time

Our data indicate that movement time increased as the difficulty of tasks increased. This, of course, is the basic premise of Fitts' law. On each of the three display size conditions, participant's movement time increased with ID ($p < .001$).

From Figure 8 we can identify some interesting effects of display size. Here we observe that participants were consistently quicker on the medium display when compared to the other two display conditions.

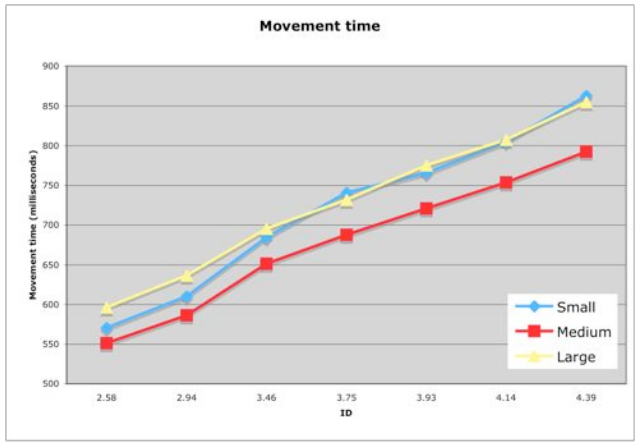

**Figure 8**. Mean movement time per ID and display conditions.

To clearly visualize this observation, we present our data grouped by display condition. In Figure 9 we see that participants were quickest on the medium display ($p < .001$).

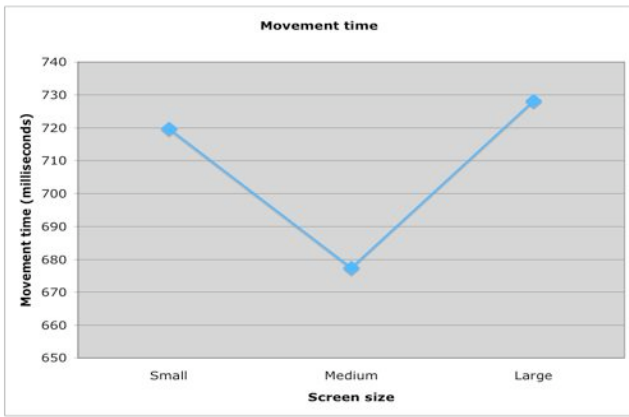

**Figure 9**. Mean movement time per display condition.

Figure 9 depicts the U shape function identified in the ergonomics literature [e.g. Hess 1973], and which has subsequently been identified in mouse-pointer studies [e.g. Accot & Zhai, 2001; Jellinek & Card, 1990], in relation to amplitude. The basic principle of the U shape function is that performance is optimized on medium-sized controls, while for larger or smaller scales performance drops. This would suggest that people perform better on a medium sized display.

An additional factor we must consider is that our participants' experience was mostly confined to standard desktop-sized displays similar to the medium sized display of our experiment. This suggests an explanation for the better performance with the medium display.

We propose that our participants, being used to medium sized displays, have adapted to a specific ratio between Motor Space and Display Space (M-D ratio). This ratio translates hand movement to cursor movement in relation to the physical environment (Figure 10). When pointer sensitivity or pixel size change, so does the M-D ratio. To keep an M-D ratio constant, the pointer sensitivity and pixel size lines in Figure 10 must move in *opposite directions* and by *proportional amounts*. A change in the pixel size line by *d* degrees must be accompanied by a change in the pointer sensitivity line of *d* x *MD* degrees in the opposite direction.

In our experiment, the use of a much smaller and much larger display disturbed the M-D ratio that participants were used to. This happened because the changes in pixel size were not coupled with a proportional change in pointer sensitivity, since we kept the latter constant.

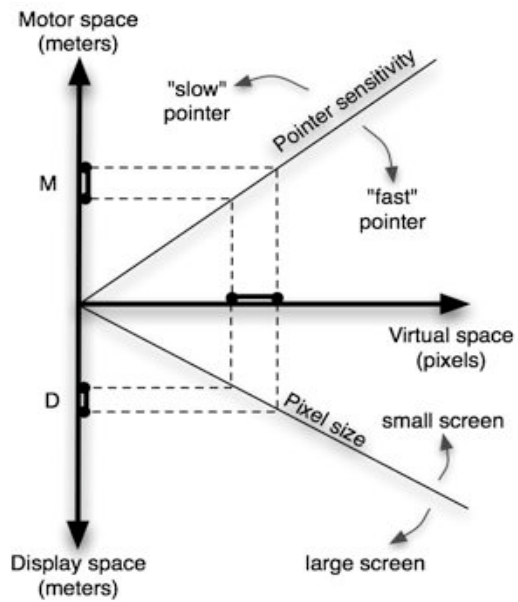

**Figure 10**. The Motor Space-Display Space (M-D) ratio describes the relationship between motor space movement (M) and display space movement (D).

Considering our experimental conditions, moving from the medium display to the small display reduced the M-D by a factor of 2.1. This meant that 2.1 times more motor space was required to cover the same display space. In this case, the pointer would need to have been made more sensitive to keep the M-D ratio constant. A graphical representation of this effect can be seen in Figure 10: as the pixel size line moves up, the pointer sensitivity line must move down in order to keep the M-D ratio constant.

Similarly, moving from the medium display to the large display increased the M-D ratio by a factor or 2.8. This meant that the same motor space covered 2.8 times more display space. In this case, pointer sensitivity would need to decrease to keep the M-D ratio constant. The graphical representation can be seen in Figure 10: as the pixel size line moves down, the pointer sensitivity line must move up in order to keep the M-D ratio constant.

The deviation from the familiar desktop-bound M-D ratio in both the small and large display conditions made it more difficult for participants, who had to take more time to complete the tasks. Additionally, the difference in movement time between small and large displays can be explained in terms of these deviations in the M-D ratio. The small display condition was slightly closer to the original ratio (2.1 as opposed to 2.8 times out), which could explain why participants were slightly quicker with the small display than with the large display.

## 6.2 Errors

Although Fitts' law makes predictions only about movement time, we also considered errors as an indicator of performance. Looking at the types of errors in each condition supports our explanation of participants' poor performance on both the small and large displays. Although participants were slow on both the small and large displays, their performance was *qualitatively* different. To highlight this, in Figure 11 we show errors per display condition.

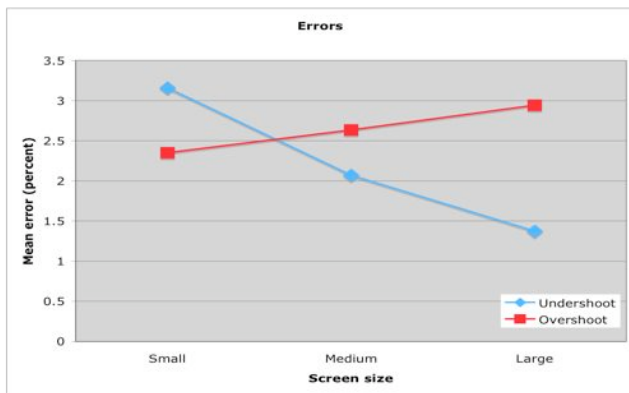

**Figure 11**. Mean undershoot and overshoot errors per display condition.

The first point to make about Figure 11 is the presence of an interaction effect. We see that the predominant type of error on the small display was undershooting, while on the large display most errors were overshoots ($p < .05$).

This observation of the effect of display size on the type of error supports the predictions of our model on the effects of increasing and decreasing pixel size. Specifically, our model predicts that *reducing pixel size results in an apparently less sensitive pointer*. Referring to Figure 10, this happens because as the pixel size line moves up, the pointer sensitivity line needs to move down in order to keep the M-D ratio constant. By keeping the pointer sensitivity line fixed, it is now *above* the position required to keep the M-D ratio constant. This was the case in our small display condition, which explains why participants had more undershoots.

Similarly, our model predicts that *an increase in pixel size results in an apparently more sensitive pointer*. This was the case in our large display condition, which led to more overshoots.

Finally, in Figure 11 we can identify that the constant pointer sensitivity we used throughout our experiment was optimized for a display size larger than 8.4" (our small display) and smaller than 18.1" (our medium display). This is the point at which the number of overshoots equals the number of undershoots.

## 6.3 The effect of amplitude on performance

Referring back to Fitts' equation, as well as Table 1 (target sizes), we see that we can keep ID constant by changing amplitude and width by the same factor. This should not have an effect on movement time since, according to Fitts' law, movement time depends only on ID. However, our results (Figure 12) suggest that amplitude does have an effect on all display sizes ($p < .001$).

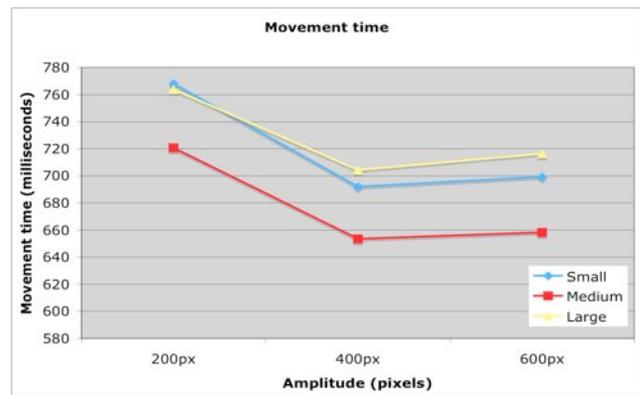

**Figure 12**. Mean movement time per amplitude and display conditions.

The 200 pixel amplitude proved most difficult ($p < .001$), for participants whilst the 400 pixel amplitude was slightly easier ($p < .05$) than the 600 pixel amplitude. We also observe that amplitude had a similar effect on each of the three display conditions. Figure 12 suggests that scale in terms of amplitude has an independent effect from scale in terms of display size. The effect of amplitude is evident by the fact that the three lines in Figure 12 are not horizontal, thus pointing to an effect of amplitude. Furthermore, the three lines do not overlap, which indicates an effect of display size.

Our data, despite confirming our hypothesis that amplitude has an effect on MT, were in certain ways unexpected. Our initial assumption was that there would be an increase in MT going from 200 to 400 to 600 pixels. Our data suggest, however, that the 200 pixel amplitude was too small, beyond accurate motor control. The result was that this condition was simply too hard for our participants, who were quite slow in it. This effect has been recorded elsewhere [e.g. Accot & Zhai, 2001], and is also suggested by the overall error rates for each amplitude condition (Figure 13).

In Figure 13 we see that the 200 pixel condition has a proportionally higher number of overshoot errors, while it ranks low in undershoots. This would suggest that 200 pixels were uncomfortably small, resulting in too many overshoots. On the other hand, the 600 pixel condition generated more undershoots than overshoots, suggesting that it was uncomfortably long.

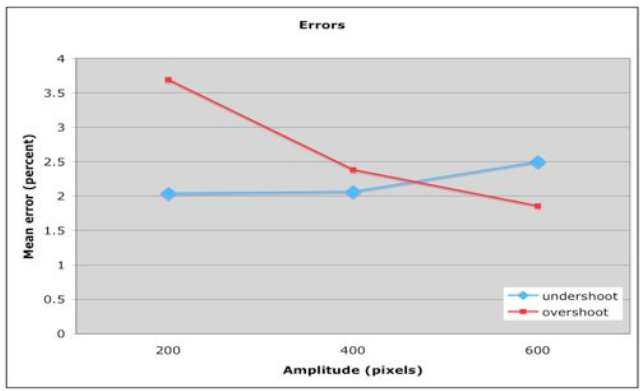

**Figure 13**. Mean error (undershoot and overshoot) per amplitude condition.

## 7  Conclusion and future work

In this paper, we set out to investigate the impact of varying display size on interaction, and specifically on target acquisition. We developed a conceptual model that relates motor space movement to display space movement via two independent parameters: pointer sensitivity and pixel size. In a study, we found that pixel size affects performance, both in terms of time and errors. Using our model we were able to explain these effects and the different types of errors (undershoots and overshoots) that we analysed.

From our analysis we are able to suggest two heuristics for users of multiple displays:

- When switching to a larger display (or lower resolution), make the mouse pointer proportionately less sensitive.

- When switching to a smaller display (or higher resolution), make the mouse pointer proportionately more sensitive.

We argue that users adapt to a specific Motor Space - Display Space ratio which, when disturbed, has a negative impact on users' performance. We do not claim that there is a universally optimum M-D ratio – rather just what each one of us is used to and adapted to. The heuristics above aim to maintain this ratio when switching between various pixel sizes, and could be implemented using software that automatically detects the size of the visual display and adjusts pointer sensitivity accordingly.

Finally, we demonstrate that scale can be considered in terms of both amplitude and pixel size, with independent effects; and our results contribute to a growing body of evidence that amplitude does indeed affect performance.

More experiments need to be conducted exploring the effects of pixel size and amplitude. The combination of different display resolutions and display sizes can yield many more possible pixel size conditions, an exploration of which is essential fully to understand the effect of pixel size on performance. Having done this, we may be able to begin formulating equations that accurately predict performance across a range of devices beyond the desktop.

## Acknowledgements

We thank Gareth Batten for all his help.

## References

ACCOT, J, AND ZHAI, S. 2001. Scale effects in steering law tasks. *Proc CHI 2001*, ACM Press, pp. 1-8.

ACCOT, J. AND ZHAI, S. 2003. Refining Fitts' law models for bivariate pointing. *Proc CHI 2003*, ACM Press, pp. 193–200.

BLANCH, R., GUIARD, Y., AND BEAUDOUIN-LAFON, M. 2004. Semantic pointing: improving target acquisition with control-display ratio adaptation. *Proc CHI 2004*, ACM Press, pp. 519-526.

FITTS, P. M. 1954. The information capacity of the human motor system in controlling the amplitude of movement. *Journal of Experimental Psychology*, 47, pp. 381–391.

GAN, K.C. AND HOFFMANN, E.R. 1988. Geometrical conditions for ballistic and visually controlled movement. *Ergonomics*, 31, pp. 829-839.

GRAWALA, M., BEERS, A.C., MCDOWALL, I., FRÖHLICH, B., BOLAS, M., AND HANRAHAN, P. 1997. The two-user responsive workbench: support for collaboration through individual views of a shared space. *Proc. SIGGRAPH 1997*, ACM Press, pp. 327-332.

GUIARD, Y. 2001. Disentangling relative from absolute amplitude in Fitts' law experiments. *Proc. CHI 2001*, Volume 2, ACM Press, pp. 315- 316.

GUIARD, Y, AND BEAUDOUIN-LAFON, M. 2004. Target acquisition in multiscale electronic worlds. *Int. J. Human-Computer Studies*, 6(16): 875-905.

HESS, R. A. 1973. Nonadjectival rating scales in human response experiments. *Human Factors*, 13(3): 275–280.

JELLINEK, H. D., AND CARD, S. K. 1990. Power mice and user performance. *Proc. CHI 1990*, pp. 213–220.

SOUKOREFF, R. W. AND MACKENZIE, I. S. 2004. Towards a standard for pointing device evaluation, perspectives on 27 years of Fitts' law research in HCI. *Int. J. Human-Computer Studies*, 61(6), pp. 751-789.B.

WEI, C. SILVA, E. KOUTSOFIOS, S. KRISHNAN, AND S. NORTH, 2000. Visualization Research with Large Displays, *IEEE Computer Graphics and Applications*, 20(4), pp. 50-54.

ZHAI, S. 2004. Characterizing computer input with Fitts' law parameters – the information and non-information aspects of pointing. *Int. J. Human-Computer Studies*, 61(6), pp. 791-809.